# A pico-Tesla magnetic sensor with PZT bimorph and permanent magnet proof mass

G. Srinivasan,[1] G. Sreenivasulu,[1] and Peng Qu[2] and Hongwei Qu[2]
[1]Physics Department and [2]E&CE
Oakland University
Rochester, MI 48309, USA

Vladimir Petrov
Institute for Information Systems
Novgorod State University
Veliky Novgorod, Russia

*Abstract*—Ferromagnetic-ferroelectric composites have attracted interests in recent years for use as magnetic field sensors. The sensing is based on magneto-electric (ME) coupling between the electric and magnetic subsystems and is mediated by mechanical strain. Such sensors for ac magnetic fields require a bias magnetic field to achieve pT-sensitivity. Here we discuss measurements and theory for a novel ac magnetic sensor that does not require a bias magnetic field and is based on a PZT bimorph with a permanent magnet proof mass. Mechanical strain on the PZT bimorph in this case is produced by interaction between the applied ac magnetic field and remnant magnetization of the permanent magnet, resulting in an induced voltage across PZT. Our studies have been performed on sensors with a bimorph of oppositely poled PZT platelets and a NdFeB permanent magnet proof mass. Magnetic floor noise N on the order of 100 pT/√Hz and 10 nT/√Hz are measured at 1 Hz and 10 Hz, respectively. When the ac magnetic field is applied at the bending resonance of ~ 40 Hz for the bimorph, the measured N ~ 700 pT/√Hz. We also discuss a theory for the magneto-electro-mechanical coupling at low frequency and bending resonance in the sensor and theoretical estimates of ME voltage coefficients are in very good agreement with the data.

## I. Introduction

A new generation of magnetic field sensors based on layered composites of ferromagnetic and piezoelectric phases has been reported in recent years [1-4]. These magneto-electric (ME) composites are capable of converting magnetic fields into electrical fields in a two-step process: magnetic field induced mechanical strain and stress induced electric field. An applied ac magnetic field H produces a magnetostrictive strain in the ferromagnetic layer, leading to an induced voltage V in the ferroelectric layer of thickness t. The ME voltage coefficient (MEVC) = V/(t H) and the ME sensitivity S=V/H are directly proportional to the product of the piezomagnetic and piezoelectric coefficients. Such ferroelectric-ferromagnetic composite sensors, however, require a bias magnetic field in order to enhance the piezomagnetic coupling coefficient in the composite to achieve pico-Tesla sensitivity [2-8]. The composite sensors have pT-sensitivity, passive, and are miniature in size and show superior performance and cost advantage over traditional flux-gate magnetometers or Hall effect sensors [3-6].

This report is on a novel pT-magnetic sensor based on ME coupling in a PZT bimorph with a permanent magnet proof mass and has the advantage of not requiring a bias magnetic field for operation. Mechanical strain on the PZT bimorph in this case is produced by interaction between the applied ac magnetic field and static remnant magnetization of the permanent magnet, resulting in an induced voltage across PZT [5]. Sensors with a bimorph of oppositely poled PZT platelets and NdFeB permanent magnet proof mass have been studied. A giant magneto-electric effect with MEVC of 20 V/cm Oe at low frequencies and enhancement to ~ 500 V/cm Oe at bending resonance have been measured for the sensor. The measured magnetic floor noise is on the order of 100 pT/√Hz to 10 nT/√Hz at 1-10 Hz. When the ac magnetic field is applied at the bending resonance for the bimorph the measured equivalent magnetic noise is ~ 700 pT/√Hz. A model is developed here and is based on equations for the strain and electric displacement of piezoelectric bimorph due to strain produced by interaction between H and remnant magnetization M. For finding the low frequency and resonance ME voltage coefficients, we solve elastostatic and electrostatic equations in PZT, taking into account boundary conditions. In this case, the theoretical modeling is similar to ME coupling in a ferromagnetic and ferroelectric bilayer [1]. The MEVC has been estimated as a function of frequency and is found to be in very good agreement with the data.

## II. Experiment

The sensor fabricated and characterized in the present study is schematically shown in Fig.1 and consisted of a cantilever of two oppositely poled piezoelectric layers of length 50 mm, width 10 mm, and thickness 0.15 mm. We used commercial PZT (No.850, APC International) platelets and were poled by heating to 400 K and cooling to room temperature in a field of 30 kV/cm. The two oppositely poled PZT platelets were bonded to each other with a 2 μm thick epoxy. The bimorph was clamped at one end and a magnet assembly of two NdFeB magnets was epoxy bonded to top and bottom of the bimorph at the free end. The magnets were circular discs of diameter 5 mm, 10 mm in height and a mass of 2.5 g each. The remnant magnetization M of NdFeB magnet assembly (along direction

Research supported by a grant from the National Science Foundation (ECCS- 1307714).

3) was measured to be 15 kG. An ac magnetic field H generated by a pair of Helmholtz coils was applied parallel to the bimorph length (direction *1*) so that interaction with M gives rise to a piezoelectric strain in PZT and a voltage V across the bimorph thickness. Since the PZT platelets are poled in opposite directions and the strain produced is compressive in one of them and tensile in the other the ME voltage in PZT layers (measured along direction *3*) will be of opposite sign so that the overall ME response is enhanced with the use of a bimorph [5].

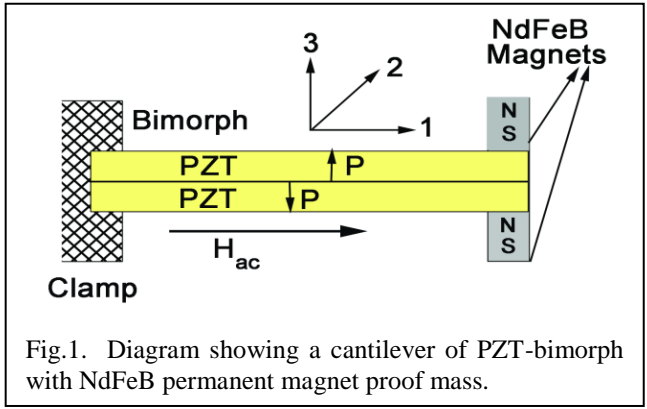

Fig.1. Diagram showing a cantilever of PZT-bimorph with NdFeB permanent magnet proof mass.

Measurements of ME sensitivity and magnetic noise were carried out by placing the sample in a plexiglass holder in magnetically shielded µ-metal chamber surrounded by an acoustic shield. The sample clamped at one end was subjected to an ac magnetic field H produced by a pair of Helmholtz coils powered by a constant current source (Keithley, model 6221). The ME voltage generated across the thickness of the bimorph was measured with a signal analyzer (Stanford Research Systems, model SR780). Since the ME voltage across PZT is nonuniform along the length of the bimorph, we measured the ME voltage V close to the clamped end where one expects maximum value [8]. The ME sensitivity S = V/H and the ME voltage coefficient MEVC = S/t (t is the bimorph thickness) were measured as a function of frequency and at room temperature. Measurements of sensor noise were performed with the signal analyzer and was converted to equivalent magnetic noise.

III. RESULTS AND DISCUSSION

A. *Magnetoelectric sensitivity*

The ME sensitivity S and MEVC were measured by measuring the voltage induced in the bimorph due to the applied ac filed H. Figure 2 shows representative results on S vs f for the specific case of H = 1 mOe at 1 Hz. The ME voltage at 1 Hz measured across the bimorph was V = 68 µV, corresponding to S = 6800 V/T and MEVC = 23 V/cm Oe. Figure 2, in addition to voltage response at 1 Hz, also shows the noise spectra for frequencies up to 14 Hz and one notices a relatively large background noise over 5-7 Hz and 9-10 Hz. Measurements of MEVC as a function of frequency showed a constant MEVC over the frequency interval f = 1-25 Hz.

Results on MEVC vs f over 25-50 Hz are shown in Fig.3. One observes frequency independent MEVC in Fig.3 except for the frequency range 33-43 Hz over which a resonance enhancement is clearly evident. The MEVC increases sharply with increasing f from 30 V/cm Oe at 30 Hz to a peak value of 480 V/cm Oe at 38 Hz. With further increase in f, the MEVC decreases rapidly and levels of at ~ 10 V/cm Oe for f > 47 Hz. The peak in MEVC is associated with bending resonance in the the bimorph with proof mass [1,5]. Similar resonances in MEVC are reported for bending modes and longitudinal and thickness acoustic resonance in ferromagnetic-piezoelectric composites [1-6]. The ME coupling at resonance is due to the traditional strain mediated coupling, but the ac field is applied at the bending mode frequency so that the overall strain and the strength of ME coupling are enhanced. Data in Figs. 2 and 3 reveal an increase in MEVC at resonance by a factor of 24 compared to low frequencies values. Thus our PZT bimorph-magnet proof mass system show a giant ME effect both at low frequencies and at bending resonance. It is noteworthy here that the permanent magnet proof mass provides an avenue for control of the resonance frequency. The bending mode frequency is found to decrease with increasing proof mass.

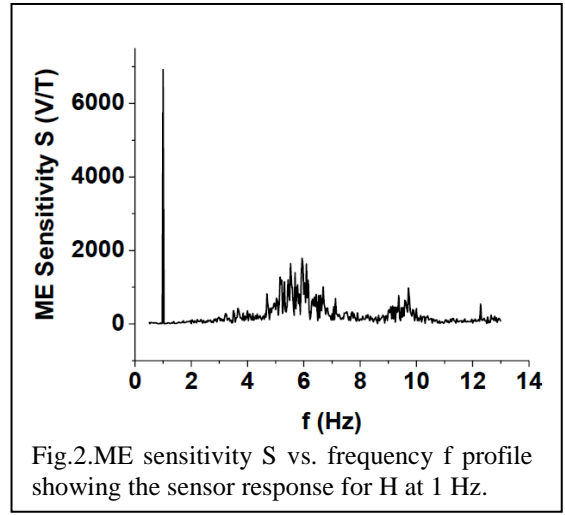

Fig.2. ME sensitivity S vs. frequency f profile showing the sensor response for H at 1 Hz.

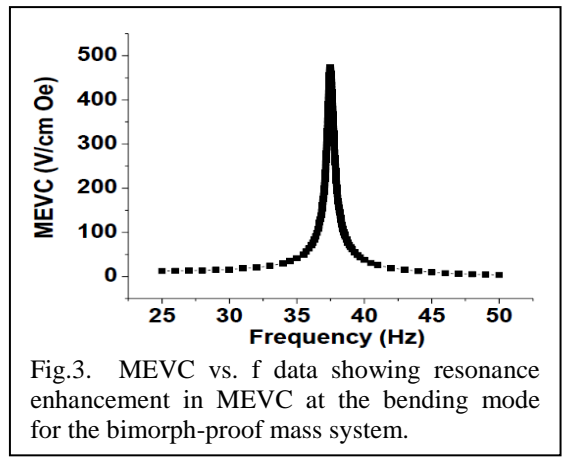

Fig.3. MEVC vs. f data showing resonance enhancement in MEVC at the bending mode for the bimorph-proof mass system.

Next we compare the results on MEVC in Fig.2 and 3 with reported values for similar systems and composites. Xing, et. al., investigated the ME coupling in PZT-bimorph loaded with permanent magnet tip mass and measured MEVC of 16 V/cmOe and 250 V/cm Oe at low frequency and bending resonance, respectively [5]. Thus the MEVC in our case are much higher than reported values in Ref.5. Past studies in the case of ferromagnetic-ferroelectric composites include ferrites, lanthanum manganites, 3-d transition metals and rare earths and alloys for the ferromagnetic phase and PZT, lead magnesium niobate-lead titanate (PMN-PT), quartz and AlN for ferroelectric/piezoelectric phase [1]. The ME sensitivity at 1 Hz in Fig. 2 is two orders of magnitude higher than reported values for bulk ferrite-piezoelectric composites and for bilayers and trilayers of ferrite-PZT and lanthanum manganite-PZT [1]. And it compares favorably with MEVC of 3 - 52 V/cm Oe at 1 kHz for Metglas composites with PZT fibers and single crystal PMN-PT [2]. The resonance MEVC in Fig.3 is higher than for ferrite based composites, but is smaller than the best value of ~ 1100 V/cm Oe reported for Metglas-PMN-PT [1, 2].

### B. Sensor noise measurements

We also measured the noise in the system for possible use as magnetic sensors. Data on equivalent magnetic noise floor were obtained over 0.5 Hz-50 Hz. The equivalent magnetic noise $N$ in terms of $T/\sqrt{Hz}$ was estimated from the measured noise (in $V/\sqrt{Hz}$) and the ME sensitivity $S$ (in $V/T$) from data in Figs.2 and 3. Results on low-frequency $N$ vs $f$ are shown in Fig.4 for our samples with PZT bimorph and magnet tip mass. One notices a general increase in $N$ from 100 pT/√Hz at 1 Hz to ~ 1 nT/√Hz at 10 Hz.

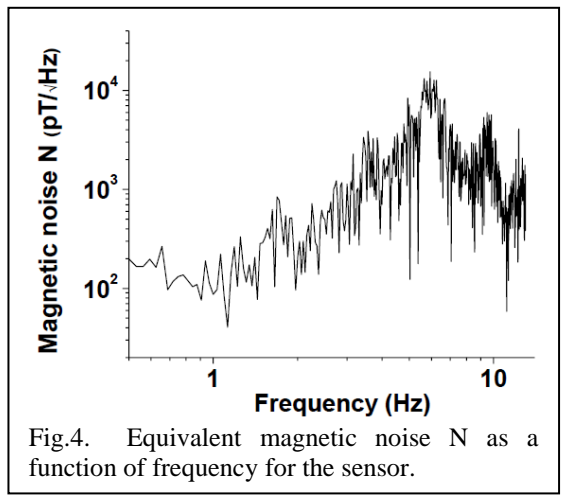

Fig.4. Equivalent magnetic noise N as a function of frequency for the sensor.

The data on noise N vs $f$ over 30-60 Hz in Fig.5 shows a constant value of N=10 nT/√Hz away from bending resonance frequency and N decreases sharply to ~ 700 pT/√Hz close to the bending mode frequency. Minor peaks of unknown origin are seen above and below the resonance frequency. Now we compare the $N$-values for our sensor with reported values for ferromagnetic-piezoelectric sensors. The best $N$ values reported to-date are for PZT or PMN-PT fibers and Metglas based sensors. Gao, et al., in their work on comparison of sensitivity and noise floor for ME sensors reported $N$ ranging from 20 to 150 pT/√Hz (at 10 Hz), respectively, for Metglas with PZT fibers or single crystal PMN-PT [9]. Wang, et al., reported a further reduction in $N$ to 5 pT/√Hz at 1 Hz for Metglas/PMN-PT fiber sensors [2]. Thus the magnetic noise for the sensor studied here compare favorably with reported values for multiferroic composite sensors.

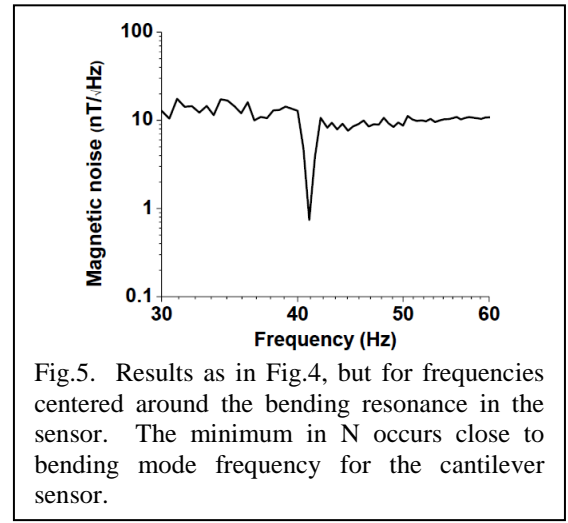

Fig.5. Results as in Fig.4, but for frequencies centered around the bending resonance in the sensor. The minimum in N occurs close to bending mode frequency for the cantilever sensor.

### C. Theory

A model for the magneto-electric response of the bimorph with permanent magnet proof mass is considered next. The specific focus is on low-frequency and voltage versus frequency characteristics around the bending resonance frequency. A cantilever with PZT layers in $(x,y)$ or $(1,2)$ plane as in Fig.1 is assumed with one end clamped and the permanent magnet assembly on the free end. The thickness of PZT along $z$-direction (direction 3) is assumed to be small compared to its length or width. The interaction between the ac magnetic field along direction 1 and remnant magnetization of the magnet along direction 3 gives rise to a piezoelectric strain in PZT. Based on equation of bending vibrations [8,10], the general expression for displacement $w$ in $z$ direction perpendicular to the sample plane can written as

$$w = C_1 \sinh(kx) + C_2 \cosh(kx) + C_3 \sin(kx) + C_4 \cos(kx), \quad (1)$$

with the wave number $k$ is defined by the expression

$$k^4 = \omega^2 \frac{2t\rho}{D}, \quad (2)$$

where $\omega$ is the angular frequency, $t$ is the thickness of each PZT layer, $\rho$ is the density, and $D$ is cylindrical stiffness of the cantilever. The constants in Eq. 1 should be determined from boundary conditions that have the following form for the cantilever with attached permanent magnet at free end:

$w=0$ and $\partial w/\partial x =0$ for $x=0$;

$A_y=\partial w/\partial x \, I \, \omega^2/b +\mu_0 JHv/b$ and $V_y=- m \, w \, \omega^2/b$ at $x=L$ (3)

where $m$, $v$, $I$, and $J$ are mass, volume, moment of inertia of magnet with respect to axis that is positioned in the middle plane along $y$ axis, and remanent magnetization, respectively, $A_y$ is the torque moment relative to $y$-axis produced by internal stresses in the bimorph per unite width, $V_y$ is the transverse force per unite width, $H$ is applied ac magnetic field, and $L$ is the sample length and $b$ is its width.

The induced electric field in PZT can be found from the open circuit condition $\int_G {}^{1,2}D_3 dx = 0$ where ${}^{1,2}D_3$ is electric induction in first and second piezoelectric layers, $G$ is the cross-section of sample normal to the $z$-axis, and ${}^{1,2}D_3 = \pm d_{31}{}^{1,2}T_1 + \varepsilon_{33}{}^{1,2}E_3$. Here $d_{31}$ and $\varepsilon_{33}$ are piezoelectric coupling coefficient and permittivity of PZT and ${}^{1,2}E$ is internal electric field in layers. The stress components ${}^{1,2}T$ can be expressed in terms of strain components ${}^{1,2}S$ from elasticity equations ${}^{1,2}T_1 = Y({}^{1,2}S_1 - d_{31}{}^{1,2}E_3)$ where Y is the modulus of elasticity of piezoelectric component at constant $E$ and ${}^{1,2}S_1 = -z_{1,2}\frac{\partial^2 w}{\partial x^2}$ ($z_{1,2}$ is distance of current point of first or second layer from the middle plane). One obtains the following equation for the low frequency MEVC for the condition that the bimorph length is much higher than the width or thickness:

$$MEVC = \frac{3v\mu_0 Jd_{31}}{4t^2 b\varepsilon_{33}}, \quad (4)$$

where $d_{31}$ is the piezoelectric coupling coefficient, and $\mu_0$ and $\varepsilon_{33}$ are the permeability and permittivity, respectively.

The influence of the magnet mass on resonance frequencies and MEVC is specified by the ratio of tip mass $m$ to bilayer mass $m_0$ and the dependence of $m$ vanishes when the mass is much than bimorph mass. A significant decrease in the bending mode frequency is expected when the tip mass is of the order of bilayer mass. The fundamental bending mode frequency is given by

$$f_r = \frac{t}{\pi L^2}\sqrt{\frac{3Y}{\rho\left(12\,m/m_0 + 3\right)}}. \quad (5)$$

The peak ME voltage coefficient at bending resonance frequency is estimated to be

$$MEVCr = \frac{9Qv\mu_0 Jd_{31}\left(5\,m/m_0 + 1\right)}{40\varepsilon_{33}bt^2\left(4\,m/m_0 + 1\right)} \quad (6)$$

with Q denoting the quality factor for bending resonance. We applied the model to estimate the ME voltage coefficient for the piezoelectric bimorph cantilever with attached magnet at free end. Resonance losses are taken into account by a using a complex frequency $\omega+i\omega'$ with $\omega'/\omega=1/Q$, and Q was estimated from observed resonance profiles. The following material parameters were used for the calculations: $Y=0.65\cdot10^{11}$ N/m$^2$, density of PZT $\rho=7.7\cdot10^3$ kg/m$^3$, $d_{31}=-1750\cdot10^{-12}$ m/V, $\varepsilon_{33}/\varepsilon_0=1750$, $m=5$ g and $\mu_0 J=1.5$ T.

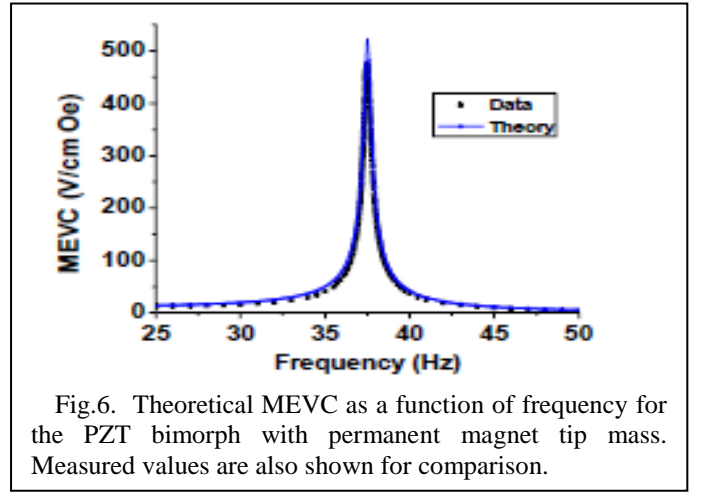

Fig.6. Theoretical MEVC as a function of frequency for the PZT bimorph with permanent magnet tip mass. Measured values are also shown for comparison.

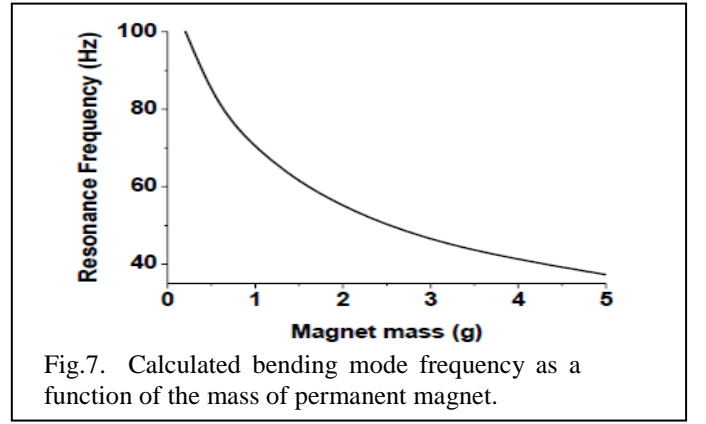

Fig.7. Calculated bending mode frequency as a function of the mass of permanent magnet.

Theoretical estimates of MEVC vs. frequency are shown in Fig. 6. Measured values in Fig.3 are also shown for comparison. One observes a very good agreement between theoretical MEVC vs f profile and the data. Both the values of MEVC and the bending mode frequency are within 2% of the measured value. Calculated values of the bending mode frequency are plotted as a function of the mass of the permanent magnet in Fig. 7. One infers the following from the results in Fig.7: (i) The cantilever arrangement facilitates electromechanical resonance at low-frequencies compared to longitudinal or thickness acoustic modes; (ii) it is possible to control the resonance frequency with proper choice for the mass of the permanent magnet; and (iii) assuming a linear increase in M with the magnet mass, any decrease in the resonance frequency with increasing m will be accompanied by an increase in the peak MEVC.

## IV. CONCLUSION

A sensor of ac magnetic fields consisting of a PZT bimorph with a permanent magnet for proof mass has been designed and characterized. The sensor operation is based on magneto-electric interaction mediated by mechanical strain. The applied ac magnetic field interacts with the remnant magnetization of the permanent magnet resulting in a strain that gives rise to a voltage response from the PZT bimorph. Magneto-electric characterization of the sensor clamped at one end indicate a giant ME effect both at low-frequencies and at bending resonance and the MEVC are comparable to best values reported for ferromagnetic and ferroelectric composites. The equivalent magnetic noise range from 100 pT/√Hz to 10 nT/√Hz , depending on the frequency. A model for the sensor has been developed and estimates of low frequency and resonance MEVC are in very good agreement with the data. The key advantages of the sensor are (i) the elimination of the need for a dc magnetic bias field that is required for high sensitivity ferromagnetic—ferroeleectric magnetic sensors and (ii) potential for control of the sensitivity by controlling the MEVC and bending resonance frequency with proper choice of proof mass. It is possible to decrease the resonance frequency and increase MEVC by increasing the proof mass so that high sensitivity could be achieved by operating the sensor under frequency modulation [11].

## ACKNOWLEDGMENT

The research was supported by a grant from the National Science Foundation (ECCS-1307714).